\documentstyle[prl,epsfig,aps,multicol]{revtex}

\begin{document}
\draft

\title{Tricritical point of lattice QCD with Wilson quarks at finite temperature and density}

\author{Xiang-Qian Luo\thanks{Email address: stslxq@zsu.edu.cn}}
\address{
CCAST (World Laboratory), P.O. Box 8730, 
Beijing 100080, China \\
Department of Physics, Zhongshan (Sun Yat-Sen) University, 
Guangzhou 510275, China\thanks{Mailing Address.} 
}

\date{\today}

\maketitle

\begin{abstract}
First principle study of QCD at finite temperature $T$ and chemical potential $\mu$
is essential for
understanding a wide range of phenomena from heavy-ion collisions to  cosmology and neutron stars.
However, in the presence of finite density,
the critical behavior lattice gauge theory without species doubling,  
is  unknown. 
At strong coupling, we examine the phase structure on the $(\mu,T)$ plane, using 
Hamiltonian lattice QCD with Wilson fermions.
A tricritical point is found, separating the
first and second order chiral phase transitions. Such a tricritical point at finite $T$ 
has not been found in previous work  
in the Hamiltonian formalism with Kogut-Susskind fermions or naive fermions.
\end{abstract}

\pacs{12.38.Gc, 11.10.Wx, 11.15.Ha, 12.38.Mh
\\ Published in Phys. Rev. D {\bf 70}, 091504 (Rapid Commun.) (2004).}

\begin{multicols}{2}

\section{Introduction}

One of the most challenging issues in particle physics is to study QCD
in extreme conditions.
Precise determination of the QCD phase diagram on temperature $T$ and chemical
potential $\mu$ plane will provide valuable information for quark-gluon
plasma (QGP) and neutron star physics. 

For QCD with two massless quarks, 
several approximation models 
(e.g. linear sigma model, Nambu-Jona-Lasino model, random matrix model,
statistical bootstrap)\cite{Halasz:1998qr}
suggest the existence of a tricritical point on the $(\mu,T)$ plane
separating the first order transition line at lower $T$ and larger $\mu$,
and the second order transition line at higher $T$ and smaller $\mu$.
There has been a proposal\cite{Stephanov:1998dy}
 for experimental search for the tricritical point,
via event-by-event fluctuations in heavy-ion collisions.

Lattice gauge theory (LGT), 
proposed by Wilson\cite{Wilson:1974sk}
is the most reliable non-perturbative approach to QCD,  based on first principles. 
Unfortunately, it suffers problems like the complex action at finite $\mu$ 
and species doubling with naive fermions.

Kogut-Susskind (KS)'s approach to lattice fermions thins the degrees of freedom, 
and preserves the remnant of chiral symmetry,
but it doesn't completely solve the species doubling problem; 
It breaks the flavor symmetry as well. Wilson's approach to lattice fermions\cite{Wilson:1974sk}
has been extensively used in hadron spectrum calculations 
as well as
in QCD at finite temperature;
It avoids the species doubling and preserves the flavor symmetry,
but it explicitly breaks the chiral symmetry, 
one of the most important symmetries  of the original theory;
Non-perturbative fine-tuning of the bare fermion mass has to be done,
in order to define the chiral limit.

In Lagrangian formulation of SU(3) LGT at finite $\mu$, 
complex action spoils numerical simulations with importance sampling.
The recent years have seen enormous efforts\cite{Muroya:2003qs,Katz:2003up,Lombardo:2004uy} 
on solving the complex action problem,
and some very interesting information\cite{Fodor:2002hs} on the phase diagram for
QCD with KS fermions at large $T$ and small $\mu$
has been obtained.  Nevertheless, to precisely locate 
the tricritical point and critical line at large $\mu$ is still an extremely difficult task.
One has also to resolve the contradiction between Monte Carlo (MC) simulations at intermediate coupling 
and strong coupling analysis
for 1 staggered flavor (corresponding to 4 flavors in the continuum): MC data\cite{Lombardo:2004uy}
indicate that the chiral phase transition at $\mu=0$ and some finite $T_C$ 
is of first order, while strong coupling analysis\cite{Damgaard:1984ag,Nishida:2003fb} favors the second order transition;
The author of  Ref. \cite{Nishida:2003fb} 
studies the phase diagram on the $(\mu,T)$ plane, and discovers a tricritical point, 
while in MC simulation\cite{Lombardo:2004uy}, 
only a line of first order phase transition is found.  
It might well be that they belong to different universality classes.

QCD at large $\mu$ is of particular importance for neutron star or quark star physics.
Hamiltonian formulation of LGT doesn't encounter the notorious ``complex action problem''.
Recently, we proposed a Hamiltonian approach to LGT with naive fermions
at finite $\mu$\cite{Gregory:1999pm,Luo:2000xi}, and extended it to Wilson fermions 
\cite{Fang:2002rk}. The chiral phase transition at $T=0$ and some finite $\mu_C$ was found to be of first order. 
In Refs. \cite{Moreo:1986rx,LeYaouanc:1987ff}, the authors studied the phase diagram of QCD with KS fermions for 2 and 4 flavors and naive fermions for 4 flavors. 
A line of second order phase transition was found in both cases,  but
no tricritical point was found at any finite $T$.

In this paper, we study the phase diagram using Hamiltonian lattice QCD with Wilson fermions.
At strong coupling and in the non-perturbatively defined chiral limit, 
we find a tricritical point on the $(\mu,T)$ plane.
The rest of the paper is organized as follows. In Sec.\ref{our approach}, 
we derive the effective Hamiltonian at finite $T$ and $\mu$. 
In Sec.\ref{critical},
we analyze the QCD phase diagram. 
The results are summarized  in Sec.\ref{discussion}.

\section{Effective Hamiltonian at the strong coupling}
\label{our approach}

\subsection{The $\mu=0$ case}
\label{zero_potential}

We begin with QCD Hamiltonian
with Wilson fermions at $\mu=0$ 
on 1 dimensional continuum time and $d=3$ dimensional spatial discretized lattice, 
\begin{eqnarray}
H &=& M \sum_{x} \bar{\psi}(x)\psi(x)
\nonumber \\
&+& {1 \over 2a} \sum_{x}\sum_{k=\pm1}^{\pm d}\bar{\psi}(x) \left(\gamma_k-r\right)U(x,k)\psi(x+{\hat k})
\nonumber \\
&+& {g^{2} \over 2a} 
\sum_{x}\sum_{j=1}^{d} E^{\alpha}_{j}(x)E^{\alpha}_{j}(x)
\nonumber \\
&-&{1 \over ag^{2}} \sum_{p} {\rm Tr} \left(U_{p}+U_{p}^{+}-2\right),
\label{first}
\end{eqnarray}
where 
\begin{eqnarray}
M=m+ {rd \over a},
\end{eqnarray}
$m$, $a$, $r$ and $g$ are respectively the bare fermion mass, 
spatial lattice spacing, 
Wilson parameter, and bare coupling constant. 
$U(x,k)$ is the gauge link variable at site $x$ and direction ${\hat k}$.
The convention $\gamma_{-k}=-\gamma_{k}$ is used.  
$E_j^{\alpha}(x)$ is the color-electric field at site $x$ and direction $j$ and   
$U_p$ is the product of gauge link variables around an elementary spatial plaquette.

For $1/g^2 <<1$, one can integrate out the gauge fields and derive the effective Hamiltonian $H_{eff}$,
consisting of terms with two fermions and four fermions\cite{Smit:1980nf,Fang:2001ry}.

\subsection{The $T \ne 0$ and $\mu \ne 0$ case}

In the continuum, 
the grand canonical partition function of QCD at finite $T$ and $\mu$ 
is
\begin{eqnarray}
Z={\rm Tr} ~{\rm e}^{- \beta \left( H - \mu N \right)}, ~~~ 
\beta = (k_{B} T)^{-1} , 
\label{sixth}
\end{eqnarray}
where $k_B$ is the Boltzmann constant
and $N$ is particle number operator
\begin{eqnarray}
N=\sum\limits_{x} \psi^{\dagger}(x) \psi (x).
\label{seventh}
\end{eqnarray}
According to Eq.(\ref{sixth}) and following the procedure in Sec.\ref{zero_potential}, 
the role of the Hamiltonian at strong coupling is now played by
\begin{eqnarray}
H^{\mu}_{eff}=H_{eff}- \mu N .
\label{eighth}
\end{eqnarray}

The vacuum energy is the expectation value of $H-\mu N$ in its ground state $\vert \Omega \rangle$, and 
also the expectation value of $H^{\mu}_{eff}$ in its ground state $\vert \Omega_{eff} \rangle$, 
given by 
\begin{eqnarray}
E_{\Omega} &=& \langle \Omega \vert H - \mu N \vert \Omega \rangle =
\langle \Omega_{eff} \vert H ^{\mu}_{eff} \vert \Omega_{eff} \rangle .
\label{vacuum_energy}
\end{eqnarray}

\subsection{Results for large $N_c$}

As shown in Ref. \cite{Fang:2002rk}, under the mean-field approximation, i.e.,
by Wick-contracting a pair of fermion fields
in the four fermion terms in $H_{eff}$,  one can obtain
a bilinear Hamiltonian in leading order of $1/N_c$
\begin{eqnarray}
H_{eff}^{\mu} \sim H_{MFA}^{\mu}&=&A\sum_{x} \bar{\psi} (x) \psi (x) 
\nonumber\\
&+&\left( B-\mu \right) \sum_{x} \psi^{\dagger} (x) \psi (x)+C ,
\label{meanfield}
\end{eqnarray}
where
\begin{eqnarray}
A &=& M-\frac {Kd}{2aN_{c}}(1-r^2) {\bar v},
\nonumber\\
B &=& \frac {Kd (1+r^2)}{a} \left( { v^{\dagger} \over 2N_{c}} -1\right),
\nonumber\\
C &=& -\frac {Kd}{4aN_{c}} 
\left( (1+r^2) {v^{\dagger}}^2 
- (1-r^2){\bar v}^2 \right) N_sN_f.
\label{consts}
\end{eqnarray}
The coefficient $A$ plays the role of dynamical mass of quark. 
${\bar v}$ and $v^{\dagger}$
are respectively the expectation of ${\bar \psi} \psi$ and $\psi^{\dagger} \psi$ in $\vert \Omega_{eff} \rangle$, divided by
$N_s$ and $N_f$,  i.e., the total number of lattice sites and the number of flavors.
\begin{eqnarray}
K=\frac {1}{g^{2}C_N}
\end{eqnarray}
is the effective four fermion coupling constant. Here $C_N=(N_c^2-1)/(2N_c)$ is the Casimir invariant of the
SU$(N_c)$  gauge group. For $N_f=1, 2$ and $N_c \ge 3$, the next-to-leading corrections are small.

In the effective Hamiltonian (\ref{meanfield}), there are three input parameters: $r$, $m$ and $\mu$. 
Suppose we study the phase structure
of the system in the chiral limit, reached by fine-tuning the bare quark mass $m$
so that the pion becomes massless. The parameter $M$ in the chiral limit is given by\cite{Fang:2002rk}
\begin{eqnarray}
M \to M_{chiral}=-\frac {Kdr^{2}}{aN_c}{\bar v}.
\label{thirty-sixth}
\end{eqnarray}
In the chiral limit,  there are only two free parameters left: $r$ and $\mu$ in Eq. (\ref{meanfield}). 
The vacuum energy is
\begin{eqnarray}
 E_{\Omega} &=& 
2N_cN_fN_s \bigg(M_{chiral} \left(n+{\bar n} -1 \right) -\frac {Kdr^2}{a} \left(n- {\bar n} +1 \right)
\nonumber\\
 &+& \frac {Kdr^2}{a} \left(n^2 +{\bar n}^{2}+1-2{\bar n} \right) 
+\frac {Kd}{a}  \left(n+\bar {n} -2n {\bar n}-1 \right)
\nonumber\\
&-& \mu \left(n-\bar {n}+1 \right) \bigg),
\label{energy_nn}
\end{eqnarray}
where the quark number $n$ and anti-quark number ${\bar n}$
are constrained in the range of $[0,1]$ 
and determined by minimizing the grand thermodynamic potential 
\begin{eqnarray}
\Phi=E_{\Omega}-TS .
\end{eqnarray}
Here 
\begin{eqnarray}
S &=&-{2N_cN_fN_s\over a} \bigg(n \ln n+(1-n) \ln (1-n)
\nonumber\\
&+& {\bar n} \ln {\bar n}+(1-{\bar n}) \ln (1-{\bar n}) \bigg) ,
\label{entropy}
\end{eqnarray}
is the entropy.

Once $n$ and ${\bar n}$ are known, the chiral condensate and quark number density in the leading order of $1/N_c$ can directly be obtained
\begin{eqnarray}
\langle {\bar \psi} \psi \rangle &=&
{\langle \Omega \vert \sum_x {\bar \psi}(x) \psi (x) \vert \Omega \rangle \over N_f N_s} 
\to  {\bar v}  
\nonumber\\
&=&
2N_{c} (n+{\bar n} -1),
\nonumber\\
n_q &=& {\langle \Omega \vert \sum_x  \psi^{\dagger}(x) \psi (x) \vert \Omega \rangle \over 2N_c N_f N_s} -1 
\to {v^{\dagger} \over 2N_c}-1
\nonumber\\
&=&
 n-{\bar n} .
\label{seventh1}
\end{eqnarray}
According to Eqs. (\ref{consts}) and (\ref{thirty-sixth}),  in the chiral limit,
the dynamical mass of quark  $m_{dyn}=A$ is proportional to ${\bar v}$ or the chiral condensate.
For convenience, we rescale the chemical potential and temperature as
$\mu'=\mu /(Kd /a)$ and $T'=T/(Kd/a)$.

\begin{figure}[htb]
\begingroup%
  \makeatletter%
  \newcommand{\GNUPLOTspecial}{%
    \@sanitize\catcode`\%=14\relax\special}%
  \setlength{\unitlength}{0.1bp}%
\begin{picture}(2519,1943)(0,0)%
\special{psfile=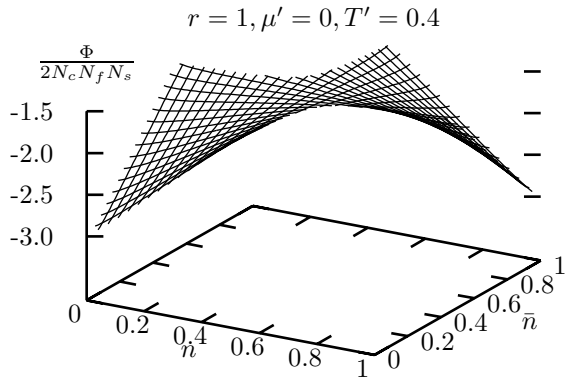 llx=0 lly=0 urx=504 ury=454 rwi=5040}
\put(403,1430){\makebox(0,0){${\Phi \over 2N_cN_fN_s}$}}%
\put(277,1252){\makebox(0,0)[r]{-1.5}}%
\put(277,1094){\makebox(0,0)[r]{-2.0}}%
\put(277,937){\makebox(0,0)[r]{-2.5}}%
\put(277,780){\makebox(0,0)[r]{-3.0}}%
\put(2072,460){\makebox(0,0){${\bar n}$}}%
\put(2163,670){\makebox(0,0)[l]{1}}%
\put(2038,599){\makebox(0,0)[l]{0.8}}%
\put(1913,528){\makebox(0,0)[l]{0.6}}%
\put(1787,457){\makebox(0,0)[l]{0.4}}%
\put(1662,385){\makebox(0,0)[l]{0.2}}%
\put(1537,314){\makebox(0,0)[l]{0}}%
\put(789,347){\makebox(0,0){$n$}}%
\put(1444,283){\makebox(0,0){1}}%
\put(1227,324){\makebox(0,0){0.8}}%
\put(1010,365){\makebox(0,0){0.6}}%
\put(793,407){\makebox(0,0){0.4}}%
\put(576,448){\makebox(0,0){0.2}}%
\put(359,489){\makebox(0,0){0}}%
\put(1259,1600){\makebox(0,0){$r=1, \mu'=0, T'=0.4$}}%
\end{picture}%
\endgroup
 
\vspace{-0.5cm}
\caption{3D plot of the grand thermodynamic potential as a function of $n$ and ${\bar n}$ 
at $r=1$, $\mu'=0$, and $T'=0.4$, where the system is in the chiral-broken phase.}
\label{fig1}
\end{figure}

\begin{figure}[htb]
\begingroup%
  \makeatletter%
  \newcommand{\GNUPLOTspecial}{%
    \@sanitize\catcode`\%=14\relax\special}%
  \setlength{\unitlength}{0.1bp}%
\begin{picture}(2519,1943)(0,0)%
\special{psfile=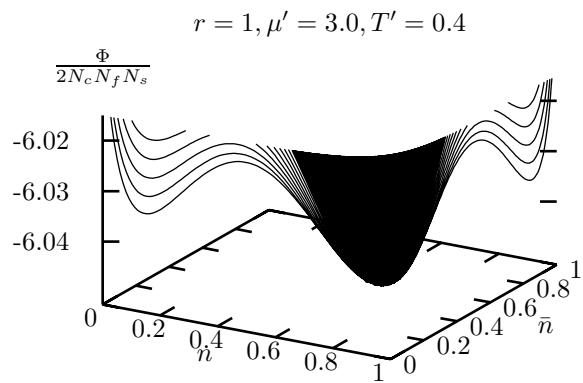 llx=0 lly=0 urx=504 ury=454 rwi=5040}
\put(403,1430){\makebox(0,0){${\Phi \over 2N_cN_fN_s}$}}%
\put(277,1157){\makebox(0,0)[r]{-6.02}}%
\put(277,966){\makebox(0,0)[r]{-6.03}}%
\put(277,776){\makebox(0,0)[r]{-6.04}}%
\put(2072,460){\makebox(0,0){${\bar n}$}}%
\put(2163,670){\makebox(0,0)[l]{1}}%
\put(2038,599){\makebox(0,0)[l]{0.8}}%
\put(1913,528){\makebox(0,0)[l]{0.6}}%
\put(1787,457){\makebox(0,0)[l]{0.4}}%
\put(1662,385){\makebox(0,0)[l]{0.2}}%
\put(1537,314){\makebox(0,0)[l]{0}}%
\put(789,347){\makebox(0,0){$n$}}%
\put(1444,283){\makebox(0,0){1}}%
\put(1227,324){\makebox(0,0){0.8}}%
\put(1010,365){\makebox(0,0){0.6}}%
\put(793,407){\makebox(0,0){0.4}}%
\put(576,448){\makebox(0,0){0.2}}%
\put(359,489){\makebox(0,0){0}}%
\put(1259,1600){\makebox(0,0){$r=1, \mu'=3.0, T'=0.4$}}%
\end{picture}%
\endgroup
 
\vspace{-0.5cm}
\caption{The same as Fig. \ref{fig1}, but for $r=1$, $\mu'=3.0$, and $T'=0.4$, where the system is at the criticality.}
\label{fig2}
\end{figure}

\begin{figure}[htb]
\begingroup%
  \makeatletter%
  \newcommand{\GNUPLOTspecial}{%
    \@sanitize\catcode`\%=14\relax\special}%
  \setlength{\unitlength}{0.1bp}%
\begin{picture}(2519,1943)(0,0)%
\special{psfile=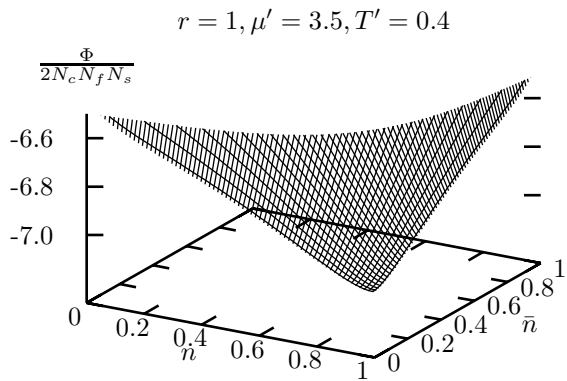 llx=0 lly=0 urx=504 ury=454 rwi=5040}
\put(403,1430){\makebox(0,0){${\Phi \over 2N_cN_fN_s}$}}%
\put(277,1160){\makebox(0,0)[r]{-6.6}}%
\put(277,977){\makebox(0,0)[r]{-6.8}}%
\put(277,795){\makebox(0,0)[r]{-7.0}}%
\put(2072,460){\makebox(0,0){${\bar n}$}}%
\put(2163,670){\makebox(0,0)[l]{1}}%
\put(2038,599){\makebox(0,0)[l]{0.8}}%
\put(1913,528){\makebox(0,0)[l]{0.6}}%
\put(1787,457){\makebox(0,0)[l]{0.4}}%
\put(1662,385){\makebox(0,0)[l]{0.2}}%
\put(1537,314){\makebox(0,0)[l]{0}}%
\put(789,347){\makebox(0,0){$n$}}%
\put(1444,283){\makebox(0,0){1}}%
\put(1227,324){\makebox(0,0){0.8}}%
\put(1010,365){\makebox(0,0){0.6}}%
\put(793,407){\makebox(0,0){0.4}}%
\put(576,448){\makebox(0,0){0.2}}%
\put(359,489){\makebox(0,0){0}}%
\put(1259,1600){\makebox(0,0){$r=1, \mu'=3.5, T'=0.4$}}%
\end{picture}%
\endgroup
 
\vspace{-0.5cm}
\caption{The same as Fig. \ref{fig1},  but for $r=1$, $\mu'=3.5$, and $T'=0.4$, 
where the system is in the chiral-symmetric phase.}
\label{fig3}
\end{figure}

Figures \ref{fig1}, \ref{fig2} and \ref{fig3} show the 3D plot of $\Phi$ as a function of
$n$ and ${\bar n}$ for $r=1$ and some $(\mu',T')$.
At $(\mu',T')=(0,0.4)$, Fig. \ref{fig1} shows there are two degenerate global minima in $\Phi$ at
$n={\bar n}=0$ and $n={\bar n}=1$; they give respectively $\langle {\bar \psi} \psi \rangle/ (2N_c)=-1$ 
and 1, and change from one to another through a chiral transformation or $(n,{\bar n}) \to (1-{\bar n},1-n)$.
At $(\mu',T')=(3,0.4)$, Fig. \ref{fig2} tells there are three degenerate global minima in $\Phi$ at
$(n,{\bar n})=(0.1448,0)$,  $(1,0.8552)$,  and $(0.9519,0.0481)$;
they lead to  $\langle {\bar \psi} \psi \rangle/ (2N_c)=-0.8552$, 0.8552, and 0 respectively. 
At $(\mu',T')=(3.5,0.4)$, Fig. \ref{fig3} indicates there is only one global minimum in $\Phi$ 
at $(n,{\bar n})=(0.9809,0.0191)$; 
this leads to $\langle {\bar \psi} \psi \rangle/ (2N_c)=0$.

\section{Phase structure on the $(\mu,T)$ plane}
\label{critical}

\begin{figure}[htb]
\begingroup%
  \makeatletter%
  \newcommand{\GNUPLOTspecial}{%
    \@sanitize\catcode`\%=14\relax\special}%
  \setlength{\unitlength}{0.1bp}%
\begin{picture}(2339,1943)(0,0)%
\special{psfile=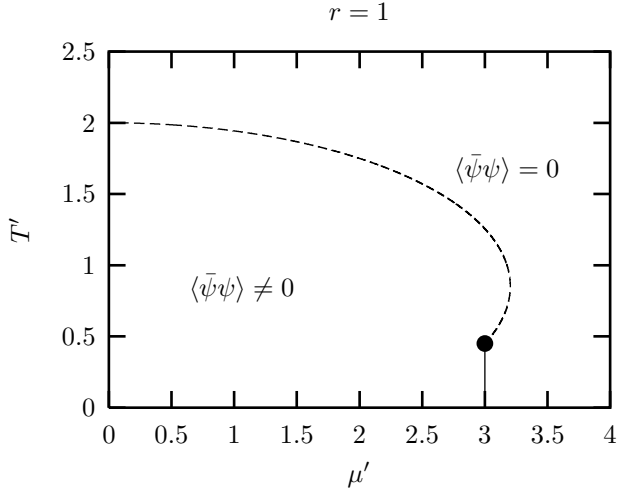 llx=0 lly=0 urx=468 ury=454 rwi=4680}
\put(1344,1793){\makebox(0,0){$r=1$}}%
\put(1344,50){\makebox(0,0){$\mu'$}}%
\put(900,750){\makebox(0,0){$\langle {\bar \psi} \psi \rangle \ne 0$}}
\put(1900,1200){\makebox(0,0){$\langle {\bar \psi} \psi \rangle = 0$}}
\put(100,971){%
\special{ps: gsave currentpoint currentpoint translate
270 rotate neg exch neg exch translate}%
\makebox(0,0)[b]{\shortstack{$T'$}}%
\special{ps: currentpoint grestore moveto}%
}%
\put(2289,200){\makebox(0,0){4}}%
\put(2053,200){\makebox(0,0){3.5}}%
\put(1817,200){\makebox(0,0){3}}%
\put(1581,200){\makebox(0,0){2.5}}%
\put(1345,200){\makebox(0,0){2}}%
\put(1108,200){\makebox(0,0){1.5}}%
\put(872,200){\makebox(0,0){1}}%
\put(636,200){\makebox(0,0){0.5}}%
\put(400,200){\makebox(0,0){0}}%
\put(350,1643){\makebox(0,0)[r]{2.5}}%
\put(350,1374){\makebox(0,0)[r]{2}}%
\put(350,1106){\makebox(0,0)[r]{1.5}}%
\put(350,837){\makebox(0,0)[r]{1}}%
\put(350,569){\makebox(0,0)[r]{0.5}}%
\put(350,300){\makebox(0,0)[r]{0}}%
\end{picture}%
\endgroup
 
\hspace{0.1cm}
\caption{Phase diagram. The solid  and dotted lines stand respectively for the first and second order transitions. 
The circle is the tricritical point.}
\label{fig4}
\end{figure}

At the second order phase transition where the chiral condensate and
the dynamical mass of quark vanish continuously, there is only one global minimum in $\Phi$,
where $\partial \Phi/\partial n = \partial \Phi /\partial {\bar n} =0$. 
Using these conditions, we obtain an equation for the critical line
\begin{eqnarray}
\mu_C' &=& \left( 1+r^2 \right) \sqrt{ 1-{2T_C' \over 1+3r^2}}  
+T_C' \ln { 1+\sqrt{ 1-{2T_C' \over 1+3r^2}} \over 
                  1-\sqrt{ 1-{2T_C' \over 1+3r^2}}} ,
\nonumber\\
\label{second_order}
\end{eqnarray}
which is depicted by the dotted line for $r=1$ in Fig. \ref{fig4}.
In the lower and left corner,
the chiral condensate and dynamical mass of  quark are non-zero.
In the other side, they vanish identically.

Below some finite $T'_3$, the situation is different. 
There is a first order chiral phase transition line 
\begin{eqnarray}
\mu'_C =1+2r^2 .
\label{firstorder}
\end{eqnarray}
from some finite $T'_3$ down to $T'=0$.
this is illustrated by the solid line 
for $r=1$ in Fig. \ref{fig4}. Chiral condensate and dynamical mass of quark
jump from non-zero for $\mu'<\mu'_C$ to zero for $\mu'>\mu'_C$.
This is consistent with the data obtained by minimizing $\Phi$, as shown in Figs. \ref{fig1}, \ref{fig2} and \ref{fig3}.
$\mu'_C$ is larger than the dynamical mass of fermion at zero
temperature. The reason, as explained in details in Ref. \cite{Fang:2002rk}, is due to
the fact that Wilson fermions break explicitly the chiral symmetry.
This seems counter-intuitive, since from a thermodynamical point of view,
a transition is expected when the chemical potential equals the dynamical
fermion mass. 
It would be interesting to see whether the difference disappears in the continuum limit. 

The points when two lines described by Eq. (\ref{second_order}) and Eq. (\ref{firstorder}) 
join are $(\mu'_3,T'_3)$ and $(\mu'_3,T'')$. $(\mu'_3,T'_3)$ is the tricritical point, while
the other one at high temperature is just a critical point on the second order phase transition line. 
Table \ref{tab1} gives the location of the tricritical point for various $r$.
For $r=1$,
we find $(\mu'_3,T'_3)=(3, 0.4498)$, i.e. the circle in Fig. \ref{fig4}. 
The phase structure for any $r\ne 0$ is qualitatively the same.
Note that  the system along the line $\mu'=\mu'_3 +0^+$ 
experiences an inverted behavior:
For $T' \in [0,T'_3)$, the system is in the chiral-symmetric phase; While for $T \in (T'_3,T'')$, 
the system enters into the chiral-broken phase; For $T' > T''$, the system is again in the chiral-symmetric phase.
The value of $T''$ is also given in Tab. \ref{tab1}. Such a behavior exists even for naive fermions, 
though there is no tricritical point at finite $T$ when $r=0$.

\begin{figure}[htb]
\begingroup%
  \makeatletter%
  \newcommand{\GNUPLOTspecial}{%
    \@sanitize\catcode`\%=14\relax\special}%
  \setlength{\unitlength}{0.1bp}%
\begin{picture}(2339,1943)(0,0)%
\special{psfile=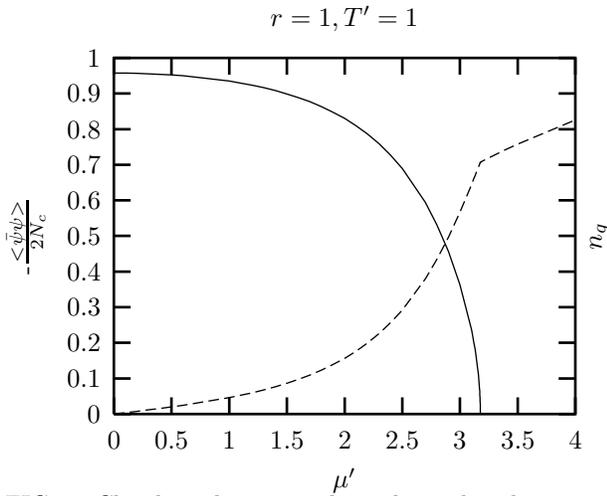 llx=0 lly=0 urx=468 ury=454 rwi=4680}
\put(1269,1793){\makebox(0,0){$r=1, T'=1$}}%
\put(1269,50){\makebox(0,0){$\mu'$}}%
\put(2238,971){%
\special{ps: gsave currentpoint currentpoint translate
270 rotate neg exch neg exch translate}%
\makebox(0,0)[b]{\shortstack{$n_q$ }}%
\special{ps: currentpoint grestore moveto}%
}%
\put(100,971){%
\special{ps: gsave currentpoint currentpoint translate
270 rotate neg exch neg exch translate}%
\makebox(0,0)[b]{\shortstack{-${<\bar {\psi} \psi> \over 2N_c}$}}%
\special{ps: currentpoint grestore moveto}%
}%
\put(2139,200){\makebox(0,0){4}}%
\put(1922,200){\makebox(0,0){3.5}}%
\put(1704,200){\makebox(0,0){3}}%
\put(1487,200){\makebox(0,0){2.5}}%
\put(1270,200){\makebox(0,0){2}}%
\put(1052,200){\makebox(0,0){1.5}}%
\put(835,200){\makebox(0,0){1}}%
\put(617,200){\makebox(0,0){0.5}}%
\put(400,200){\makebox(0,0){0}}%
\put(350,1643){\makebox(0,0)[r]{1}}%
\put(350,1509){\makebox(0,0)[r]{0.9}}%
\put(350,1374){\makebox(0,0)[r]{0.8}}%
\put(350,1240){\makebox(0,0)[r]{0.7}}%
\put(350,1106){\makebox(0,0)[r]{0.6}}%
\put(350,972){\makebox(0,0)[r]{0.5}}%
\put(350,837){\makebox(0,0)[r]{0.4}}%
\put(350,703){\makebox(0,0)[r]{0.3}}%
\put(350,569){\makebox(0,0)[r]{0.2}}%
\put(350,434){\makebox(0,0)[r]{0.1}}%
\put(350,300){\makebox(0,0)[r]{0}}%
\end{picture}%
\endgroup
 
\caption{Chiral condensate and quark number density as a function of $\mu'$ at $r=1$ and $T'=1$.
Eq. (\ref{second_order}) predicts a second order chiral phase transition at $\mu'_C=3.1770$.}
\label{fig5}
\end{figure}

\begin{figure}[htb]
\begingroup%
  \makeatletter%
  \newcommand{\GNUPLOTspecial}{%
    \@sanitize\catcode`\%=14\relax\special}%
  \setlength{\unitlength}{0.1bp}%
\begin{picture}(2339,1943)(0,0)%
\special{psfile=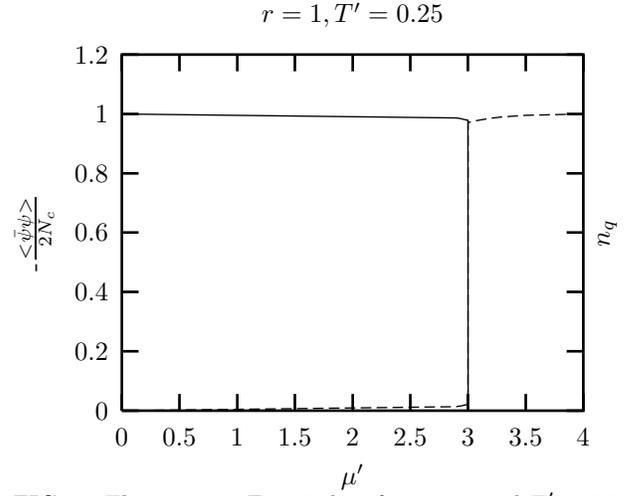 llx=0 lly=0 urx=468 ury=454 rwi=4680}
\put(1269,1793){\makebox(0,0){$r=1, T'=0.25$}}%
\put(1269,50){\makebox(0,0){$\mu'$}}%
\put(2238,971){%
\special{ps: gsave currentpoint currentpoint translate
270 rotate neg exch neg exch translate}%
\makebox(0,0)[b]{\shortstack{$n_q$ }}%
\special{ps: currentpoint grestore moveto}%
}%
\put(100,971){%
\special{ps: gsave currentpoint currentpoint translate
270 rotate neg exch neg exch translate}%
\makebox(0,0)[b]{\shortstack{-${<\bar {\psi} \psi> \over 2N_c}$}}%
\special{ps: currentpoint grestore moveto}%
}%
\put(2139,200){\makebox(0,0){4}}%
\put(1922,200){\makebox(0,0){3.5}}%
\put(1704,200){\makebox(0,0){3}}%
\put(1487,200){\makebox(0,0){2.5}}%
\put(1270,200){\makebox(0,0){2}}%
\put(1052,200){\makebox(0,0){1.5}}%
\put(835,200){\makebox(0,0){1}}%
\put(617,200){\makebox(0,0){0.5}}%
\put(400,200){\makebox(0,0){0}}%
\put(350,1643){\makebox(0,0)[r]{1.2}}%
\put(350,1419){\makebox(0,0)[r]{1}}%
\put(350,1195){\makebox(0,0)[r]{0.8}}%
\put(350,972){\makebox(0,0)[r]{0.6}}%
\put(350,748){\makebox(0,0)[r]{0.4}}%
\put(350,524){\makebox(0,0)[r]{0.2}}%
\put(350,300){\makebox(0,0)[r]{0}}%
\end{picture}%
\endgroup
 
\caption{The same as Fig. \ref{fig5}, but for $r=1$ and $T'=0.25$.
Eq. (\ref{firstorder}) predicts a first order chiral phase transition at $\mu'_C=3$.}
\label{fig6}
\end{figure}

\begin{figure}[htb]
\begingroup%
  \makeatletter%
  \newcommand{\GNUPLOTspecial}{%
    \@sanitize\catcode`\%=14\relax\special}%
  \setlength{\unitlength}{0.1bp}%
\begin{picture}(2339,1943)(0,0)%
\special{psfile=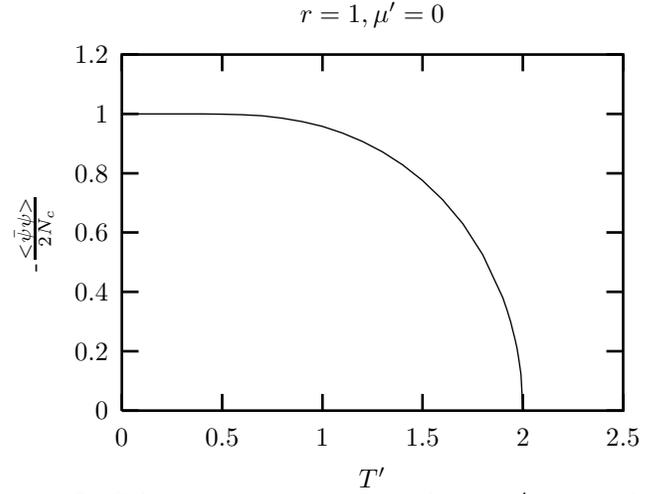 llx=0 lly=0 urx=468 ury=454 rwi=4680}
\put(1344,1793){\makebox(0,0){$r=1, \mu'=0$}}%
\put(1344,50){\makebox(0,0){$T'$}}%
\put(100,971){%
\special{ps: gsave currentpoint currentpoint translate
270 rotate neg exch neg exch translate}%
\makebox(0,0)[b]{\shortstack{-${<\bar {\psi} \psi> \over 2N_c}$}}%
\special{ps: currentpoint grestore moveto}%
}%
\put(2289,200){\makebox(0,0){2.5}}%
\put(1911,200){\makebox(0,0){2}}%
\put(1533,200){\makebox(0,0){1.5}}%
\put(1156,200){\makebox(0,0){1}}%
\put(778,200){\makebox(0,0){0.5}}%
\put(400,200){\makebox(0,0){0}}%
\put(350,1643){\makebox(0,0)[r]{1.2}}%
\put(350,1419){\makebox(0,0)[r]{1}}%
\put(350,1195){\makebox(0,0)[r]{0.8}}%
\put(350,972){\makebox(0,0)[r]{0.6}}%
\put(350,748){\makebox(0,0)[r]{0.4}}%
\put(350,524){\makebox(0,0)[r]{0.2}}%
\put(350,300){\makebox(0,0)[r]{0}}%
\end{picture}%
\endgroup
 
\caption{Chiral condensate as a function of $T'$ at $r=1$ and $\mu'=0$.
Eq. (\ref{second_order}) predicts a second order chiral phase transition at $T'_C=2$.}
\label{fig7}
\end{figure}

Figure \ref{fig5} shows the results for the chiral condensate and quark number density
(\ref{seventh1}) as a function $\mu'$ for $T'=1$ and $r=1$, above the tricritical point;
The results indicate there is a second order chiral phase transition at $\mu'=3.1770$. 
Figure \ref{fig6} shows 
those for $T'=0.25$ and $r=1$, below the tricritical point; There is a first order chiral phase transition at $\mu'=3$.
Figure \ref{fig7} shows 
the chiral condensate for $\mu'=0$ and $r=1$; There is a second order chiral phase transition at $T'=2$. These results are obtained by locating the minimum of  $\Phi$, and the critical point is consistent with the prediction of Eqs. (\ref{second_order})
and (\ref{firstorder}).

\section{Discussions}
\label{discussion}

In the preceding sections, we have investigated the QCD phase diagram in
Hamiltonian lattice formulation with Wilson fermions.
At the strong coupling, we find a
tricritical point on the $(\mu,T)$ plane, which
has not been found in previous work  
in the Hamiltonian formalism with Kogut-Susskind fermions or naive fermions.
Our findings imply that on the $(\mu,T)$ phase, the phase structure of QCD with Wilson fermions (without species doubling) might be qualitatively different from naive or Kogut-Susskind fermions (with
species doubling). 
Further detailed lattice study 
will be very important for understanding the QCD phase diagram.


\acknowledgments

I thank V. Azcoiti, Y. Fang, S. Guo, S. Katz, V. Laliena, and M. Lombardo for useful discussions.
This work is supported by the Key Project of National Science Foundation (10235040), 
and National and Guangdong Ministries of Education.

\begin{table}
\begin{center}
\begin{tabular}{|c|c|c|c|}
\hline
$r$ & $\mu'_3$ & $T'_3$ & $T''$\\ \hline 
0.0  & 1.00 & 0.0000 & 0.3407 \\
0.1  & 1.02 & 0.0016 & 0.3504\\
0.2  & 1.08 & 0.0087 & 0.3792\\
0.3  & 1.18 & 0.0238 & 0.4269\\
0.4  & 1.32 & 0.0489 & 0.4930\\
0.5  & 1.50 & 0.0853 & 0.5771\\
0.6  & 1.72 & 0.1336 & 0.6789\\
0.7  & 1.98 & 0.1942 & 0.7979\\
0.8  & 2.28 & 0.2670 & 0.9339\\
0.9  & 2.62 & 0.3523 & 1.0867\\
1.0  & 3.00 & 0.4498 & 1.2563\\
\hline 
\end{tabular}
\end{center}
\caption{Dependence of tricritical point on $r$. 
The point $T''$ described in the text is also given.}
\label{tab1}
\end{table}

\end{multicols}
\end{document}